\documentclass[aip,apl,reprint,groupedaddress]{revtex4-1}
\usepackage{epsfig,graphics}
\usepackage{graphicx}

\begin{document}

\title{Observation and characterization of mode splitting in microsphere resonators in aquatic environment}
\author{Woosung Kim}
\email{wk1@cec.wustl.edu}
\affiliation{Department of Electrical and Systems Engineering, Washington University, St. Louis, Missouri 63130, USA}
\author{\c{S}ahin Kaya \"{O}zdemir}
\email{ozdemir@ese.wustl.edu}
\affiliation{Department of Electrical and Systems Engineering, Washington University, St. Louis, Missouri 63130, USA}
\author{Jiangang Zhu}
\affiliation{Department of Electrical and Systems Engineering, Washington University, St. Louis, Missouri 63130, USA}
\author{Lan Yang}
\email{yang@ese.wustl.edu}
\affiliation{Department of Electrical and Systems Engineering, Washington University, St. Louis, Missouri 63130, USA}

\begin{abstract}
Whispering gallery mode (WGM) optical resonators utilizing resonance shift (RS) and mode splitting (MS) techniques have emerged as highly sensitive platforms for label-free detection of nano-scale objects. RS method has been demonstrated in various resonators in air and liquid. MS in microsphere resonators has not been achieved in aqueous environment up to date, despite its demonstration in microtoroid resonators. Here, we demonstrate scatterer-induced MS of WGMs in microsphere resonators in water. We determine the size range of particles that induces MS in a microsphere in water as a function of resonator mode volume and quality factor. The results are confirmed by the experimental observations.
\end{abstract}

\maketitle

Whispering gallery mode microresonators have been widely studied and used for the label-free detection of nanoparticles, viruses and pathogens \cite{Vollmer02, Vollmer01}. Resonance shift (RS) method, which is based on monitoring the spectral shift of a single resonance mode caused by the change in effective polarizability due to changes in the resonator surrounding, has been successfully applied not only for the detection of individual synthetic nanoparticles and virions, but also for the detection and characterization of biomarkers and biological phenomena in very small sample sizes. Mode splitting (MS) method, on the other hand, relies on the splitting of a single resonance mode of a WGM resonator into two by scatterer-induced coupling between two frequency-degenerate counter-propagating modes \cite{Gorodetsky01, Weiss01}, and have been recently demonstrated \cite{Zhu01} as a highly sensitive alternative to RS method. This self-reference method allows not only the detection at single particle resolution but also the single-shot measurement of particle size \cite{Zhu01}. A crucial step in effective use of RS or MS for detecting and studying biomolecules is their demonstration and characterization in various surroundings including air and aqueous environment.

Resonance shift method has been the workhorse of sensing applications of WGM resonators, and has been demonstrated in almost all existing WGM resonator geometries including microspheres, microrings and microtoroids in air, water and buffer solutions \cite{Vollmer02,Armani,Fan}. Studies of MS, however, have long been limited to microspheres and microtoroids in air, and only recently was achieved in a microtoroid in an aqueous environment \cite{Kim01}.

Up to date, demonstration of MS in microspheres and microrings in aquatic environment has not been reported, although Teraoka and Arnold have discussed the conditions under which MS can be observed in spherical resonators \cite{Teraoka01}. For the MS to become a versatile and effective method for sensing applications, its realization in different WGM resonator geometries and in various environments is very important. Here, we report nanoparticle induced MS in high-{\rm Q} microsphere resonators in aquatic environment, and investigate the parameters that affect the observation of MS in such structures.

We performed experiments using microsphere resonators with different sizes and polystyrene (PS) particles of various radii. Microspheres were fabricated by melting the tip of a tapered silica fiber with CO$_{2}$ laser. Light from a tunable laser in $660~{\rm nm}$ band is coupled into and out of the microspheres using tapered fibers. After the characterization in air, the microsphere-taper system was immersed in a deionized water droplet of $\sim50~\mu L$. PS nanoparticle samples of  $2\mu L$ was added to the droplet from a solution of concentration $\sim2\times10^{-11}\rm wt{\rm\%}$. Transmission spectra were collected continuously for $10-15 {\rm min}$. If MS was not observed, an additional $2\mu L$ from the solution was added to increase the particle concentration and hence increase the probability of MS. This was continued until MS was observed.

\begin{figure}
        \includegraphics[width=0.45\textwidth]{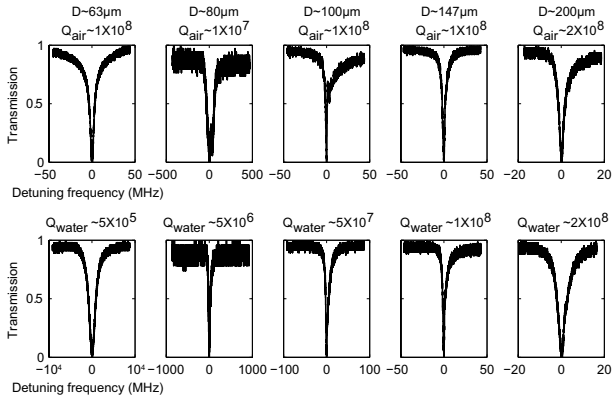}
        \caption{\footnotesize  Characterization of whispering gallery mode microsphere resonators in air and water in $660 {\rm nm}$ wavelength band. ${\rm Q_{air}}$ and ${\rm Q_{water}}$ are quality factor of the microspheres in air and in water.}\label{fig1}
\end{figure}

Transition from air to an aquatic environment shows its first effect on the resonator ${\rm Q}$, which degrades due to increased absorption losses in water. This can be partially compensated by working with resonators having resonances in the wavelength band where water absorption is minimal. For example, switching the wavelength from $1.5~{\rm \mu m}$ band to $660~{\rm nm}$ leads to more than 1000-fold decrease in absorption coefficient of light in water \cite{Hale01}. Figure \ref{fig1} shows the measured transmission spectra and the corresponding ${\rm Q}$ factors in both air and water for five microspheres of sizes $D=63-200~{\rm \mu m}$. We observe that transition from air to aquatic environment significantly affects the ${\rm Q}$ of the microspheres with smaller diameter, while the effect on larger microspheres is slight if not unobservable. For example, while the ${\rm Q}$ of the microsphere with $D=63~{\rm \mu m}$ experiences a 1000-fold decrease, the change in the ${\rm Q}$ of the microsphere with $D=100~{\rm \mu m}$ is less than 10-fold, and ${\rm Q}$ does not change significantly for the microsphere with $D=200~{\rm \mu m}$. This can be explained by the increased radiation leakage and bending loss as the size of the microsphere becomes smaller in addition to the absorption in water which affects resonators of all sizes. Note that the MS seen in Fig. \ref{fig1} for the microsphere of ${\rm D}=80 {\rm \mu m}$ in air cannot be resolved when the resonator is immersed in water.

\begin{figure}
        \includegraphics[width=0.45\textwidth]{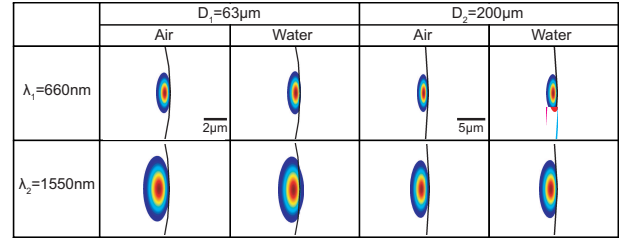}
        \caption{\footnotesize  Numerical simulation results showing the confinement of fundamental WGM modes in microspheres of different sizes in water and air at two different wavelength bands.}\label{fig2}
\end{figure}

Mode volume, which quantifies the confinement of a WGM field in the resonator and its extension in the surrounding, is another parameter affected by the transition from air to aquatic environment. Figure 2 reveals the effect of the resonator size, wavelength and the surrounding media on the confinement and the distribution of the WGM fields. Confinement in air ($n_{\rm air}=1$) is stronger than in water ($n_{\rm water}=1.33$), because refractive index contrast between the resonator ($n_{\rm silica}=1.44$) and air is larger than that between the resonator and water. It is clear in Fig. \ref{fig2} that $V_{\rm water}$ is always larger than $V_{\rm air}$; however, the difference is smaller for larger microspheres and for visible light. For example, at $\lambda=660 {\rm nm}$, mode volume of a microsphere of $D=63~(200)~{\rm \mu m}$ increases by $2.6~(0.6)\%$ in water. For the microsphere of $D=63~{\rm \mu m}$ and $\lambda=1550 {\rm nm}$, mode volume in water is $58\%$ larger than that in air, whereas for $D=200~{\rm \mu m}$ the increase is only $0.1\%$. Similarly, the value of the normalized field distribution, $f(\textbf{r})$, at microsphere surface is higher in water and its value decreases with increasing microsphere size.

The effect of transition from air to an aquatic environment on the MS experiments becomes clearer when we look at the condition to resolve MS which is denoted as $|2g|>\Gamma+\omega/{\rm Q}$. Here, $|2g|=\alpha f^{2}(\textbf{r})\omega/V$ and $\Gamma=|g|\alpha \omega^{3}/3\pi \nu^{3}$ are the amount of scatterer-induced MS and additional loss, respectively, $\omega$ is the angular resonance frequency of the WGM before the scatterer, $\rm{Q}$ is the quality factor, $\nu$ is the speed of light in the surrounding medium, and $\omega/{\rm Q}$ represents the linewidth of the pre-scatterer resonance mode. The polarizability $\alpha$ is defined as $\alpha=4\pi R^{3}(n_{p}^{2}-n_{e}^{2})/(n_{p}^{2}+2n_{e}^{2})$ for a spherical particle of radius $R$ and refractive index $n_p$ in a surrounding environment of $n_e$. Then it is clear that a spherical particle of radius $R$ has a larger polarizability in air than in water because the refractive index contrast between the surrounding and the particle is larger in air, i.e., $(n_{p}^{2}-n_{\rm air}^{2})/(n_{p}^{2}+2n_{\rm air}^2)>(n_{p}^{2}-n_{\rm water}^{2})/(n_{p}^{2}+2n_{\rm water}^2)$. Thus, provided that all the other parameters are kept constant, the smallest detectable particle size using MS method is larger in water than in air.

\begin{figure}
\includegraphics[width=0.3\textwidth]{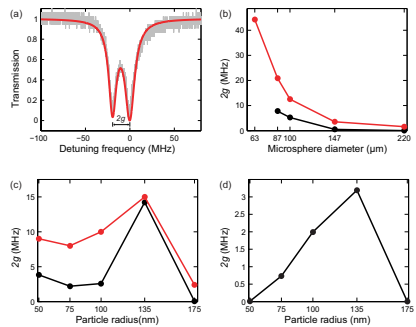}
\caption{\footnotesize  Nanoparticle-induced MS in a microsphere resonator placed in water.  (a) A typical MS spectrum observed in a microsphere of diameter ${\rm D}=87\mu{\rm m}$ for polystyrene nanoparticle of $R=50 nm$. The amount of MS is $2g=19.4{\rm MHz}$. Red line is the lorentzian fit to the experimental data. (b) $2g$ versus the size of microsphere resonator for nanoparticles of $R=50 {\rm nm}$. (c) $2g$ versus the nanoparticle size for microspheres of size ${\rm D}\sim 89 {\rm \mu m}$ (d) $2g$ versus the nanoparticle size for microspheres of size ${\rm D}\sim 214 {\rm \mu m}$. In (b) and (c), red curves are simulation results.} \label{fig3}
\end{figure}

Experimental results presented in Fig. \ref{fig3} show the dependence of $2|g|$ on the size of the microsphere and the nanoparticle. For a fixed particle size, the splitting $2|g|$ decreases as the size of the microsphere is increased. Moreover, for a fixed size of microsphere, there exists a particle size which leads to the largest $2|g|$, i.e., particles with much smaller or larger size leads to decrease in $2|g|$. Numerical simulations show a similar tendency.

For a given ${\rm Q}$ there is a threshold $R$ beyond which the MS process is dominated by scatterer-induced losses, that is $\Gamma\gg\omega/{\rm Q}$. Then $|2g|>\Gamma$ should be satisfied to ensure resolvability of MS, which, in turn, requires $\alpha<6\pi(\nu/\omega)^3$. There is an upper limit for $R$ beyond which $\Gamma$ reaches so large values (i.e., $\Gamma>2|g|$)  that MS can no longer be resolved. On the other hand, for very small $R$, cavity-related losses starts dominating, that is $\Gamma\ll\omega/{\rm Q}$. Then the condition to resolve MS becomes $|2g|>\omega/{\rm Q}$ which requires that $\alpha>V{\rm Q}^{-1}f^{-2}(\textbf{r})$. Thus, in order to detect smaller particles, one needs to assure higher ${\rm Q}/V$ and higher $f^{2}(\textbf{r})$ at the particle location. Then for a given ${\rm Q}/V$, there is a critical value of $\alpha$, hence $R$, below which MS cannot be resolved.

It is obvious that observing MS in aquatic environment is more difficult than that in air, because a particle of size $R$ has smaller polarizability, and a microsphere of size $D$ has a larger $V$ but smaller ${\rm Q}$  in water. One cannot make the $V$ arbitrarily small because this will further reduce ${\rm Q}$. There is a delicate balance between $\alpha$, ${\rm Q}$, $V$ and $f(\textbf{r})^2$ which should be satisfied to observe MS in water. Thus, detecting a nanoparticle with polarizability $\alpha$ using a microsphere resonator in aquatic environment requires careful fabrication process for the resonator such that resonators with smaller $V$ and higher ${\rm Q}$ and $f(\textbf{r})^2$ are available. In other words, the resonator should be fabricated such that $V{\rm Q}^{-1}f^{-2}(\textbf{r})$ is smaller than $\alpha$.

\begin{figure}
\includegraphics[width=0.35\textwidth]{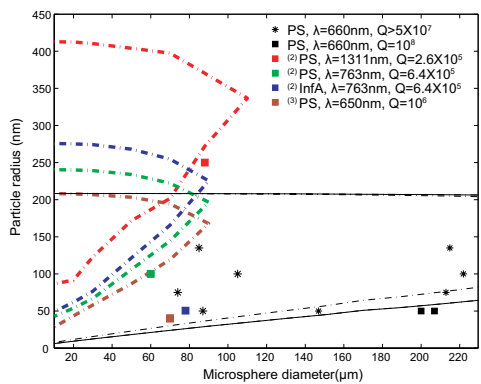} \caption{\footnotesize Constant ${\rm Q}$ contours bounding the size range of nanoparticles which can induce MS in microsphere resonators of different diameters in water. The boundaries are calculated from the MS resolvability criterion using the data, i.e., ${\rm Q}$ and $\lambda$, from our experiments (black contours, $\ast$ and black square) and those reported in Refs. \cite{Vollmer01,Shopova01} (colored contours and square boxes). MS could not be observed for the data points lying outside the boundaries denoted by the same color. The data represented by $\ast$ are the only ones which induce MS. The solid and dashed black curves denote the contours for ${\rm Q}=10^8$ and ${\rm Q}=5\times10^7$, respectively. These ${\rm Q}$ values are obtained for microspheres used in our experiments.} \label{fig4}
\end{figure}
The MS resolvability relation, $|2g|>\Gamma+\omega/{\rm Q}$, corresponds to a second order inequality in $\alpha$ whose solution is depicted in Fig. \ref{fig4}. The contours in Fig.\ref{fig4} denote the size ranges of the nanoparticles and microsphere resonators which satisfy the MS resolvability condition for a given ${\rm Q}$ of the resonator. Figure \ref{fig4} confirms the results of Fig.\ref{fig3} and above discussions showing that the range of particle polarizability (i.e., size if the refractive index is known or vice versa) for observable MS depends on the size and quality factor of the microsphere resonator. In Fig. \ref{fig4}, we also included the results of nanoparticle detection experiments reported in the literature for microsphere resonators using the RS technique. It is clearly seen that the reported ${\rm Q}$ and $D$ of the microspheres do not allow a resolvable MS for the nanoparticles tested in those works. For example, Shopova {\it et al.}\cite{Shopova01} has reported that although detection of single PS nanoparticles of radii $R=40 {\rm nm}$ has been achieved using RS technique using  a microsphere of $D=70 {\rm \mu m}$ and ${\rm Q}\sim 10^6$, MS could not be observed. Our resolvability criterion suggests that with the reported ${\rm Q}$ and $D$, one cannot observe MS with particles of $R=40 {\rm nm}$ because such particles will fall outside the detectable particle size range of  $120{\rm nm}<R<195{\rm nm}$. Thus, it is not surprising that the authors could not see MS in their experiments. Similarly, in our experiments with a microsphere of $D=200 {\rm \mu m}$ and ${\rm Q}\sim 10^8$, we could not detect PS particles of $R=50 {\rm nm}$ which falls out of the size range $57{\rm nm}<R<207 {\rm nm}$ predicted by the MS resolvability condition for the used microsphere. The same PS particle could be detected with a smaller microsphere of $D\sim 90 {\rm \mu m}$ and ${\rm Q}\sim 5\times10^7$ as it falls in the range of detectable particle size of $37{\rm nm}<R<208 {\rm nm}$. These results showing the delicate relation among microsphere size, particle size and the Q-factor of the resonator agree well with the predictions of Teraoka and Arnold \cite{Teraoka01}.
\begin{figure}
        \includegraphics[width=0.38\textwidth]{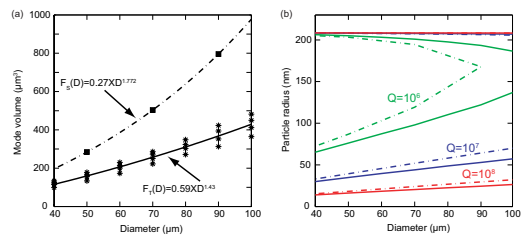}
        \caption{\footnotesize  Numerical simulation results showing the comparison of (a) mode volume and (b)MS resolvability criterion for microsphere (dashed) and microtoroid (solid) resonators in water as a function of their major diameter at $\lambda=660 {\rm nm}$ band. The $\ast$'s in (a)indicate different minor diameters ($3-6\mu{\rm m}$). In (b) constant Q contours bound the size range of particles which can induce MS.}\label{fig5}
\end{figure}

Figure \ref{fig5} clearly shows that $V_{\rm sphere}$ is at least twice as large as $V_{\rm toroid}$. This agrees well with the results of Arnold {\it et.al.} \cite{Arnold01}. Larger $V$ makes it more difficult to observe MS in microspheres. To induce MS for a given particle size, a microsphere  having the same diameter as a microtoroid requires higher $Q$. A microtoroid having the same $Q$ and the same major diameter as a microsphere can detect smaller particles.

In conclusion, we report the observation of MS in a microsphere in aquatic environment, and systematically analyze the relation between the size and quality factor of the resonator and the detectable range of particle size using the MS technique. We also discuss why nanoparticle induced MS could not be observed in earlier experiments in aqueous medium, and provide a relation to optimize the microsphere size and quality factor for the targeted range of particle sizes.

The authors acknowledge the support from MISA at Washington University in St. Louis. This work is supported by the NSF under grant No.0954941.

\end{document}